\documentclass[%
 reprint,
superscriptaddress,
 amsmath,amssymb,
 aps,
]{revtex4-2}

\usepackage{graphicx}
\usepackage{dcolumn}
\usepackage{bm}
\usepackage{braket}
\usepackage{color}
\usepackage{amsfonts}
\usepackage{hyperref}
\hypersetup{hidelinks,
	colorlinks=true,
	allcolors=black,
	pdfstartview=Fit,
	breaklinks=true
}

\makeatletter
\renewcommand{\textcolor}[2]{#2}
\makeatother

\begin{document}


\title{Memory Kernel Coupling Theory: Obtain Time Correlation Function from Higher-order Moments}

\author{Wei Liu}
  \email{These two authors contribute equally to the research}
 \affiliation{Department of Chemistry, Westlake University, Hangzhou 310030, China}
 \affiliation{Institute of Natural Sciences, Westlake Institute for Advanced Study, Hangzhou 310024 Zhejiang, China}
 
\author{Yu Su}%
  \email{These two authors contribute equally to the research}
\affiliation{ 
Hefei National Research Center for Physical Sciences at the Microscale, University of Science and Technology of China, Hefei, Anhui 230026, China
}%

\author{Yao Wang}%
 \email{wy2010@ustc.edu.cn}
\affiliation{ 
Hefei National Research Center for Physical Sciences at the Microscale, University of Science and Technology of China, Hefei, Anhui 230026, China
}%

\author{Wenjie Dou}
\email{douwenjie@westlake.edu.cn}
 \affiliation{%
Department of Chemistry, Westlake University, Hangzhou 310030, China
}%
\affiliation{Institute of Natural Sciences, Westlake Institute for Advanced Study, Hangzhou 310024 Zhejiang, China}%
\affiliation{%
Department of Physics, Westlake University, Hangzhou 310030, China
}

\date{\today}

\begin{abstract}
Dynamical observables can often be described by time correlation functions (TCFs). However, efficiently calculating TCFs for complex quantum systems is a significant challenge, which generally requires solving the full dynamics of the systems. This Letter presents the memory kernel coupling theory (MKCT), a general formalism for evaluating TCFs. The MKCT builds upon Mori's memory kernel formalism for TCFs. Our theory further decomposes the memory kernel into auxiliary kernels. Rapid decay of auxiliary kernels allows us to truncate the coupled equations of motion with high accuracy. Notably, only higher-order moments are sufficient as the input for obtaining TCFs. While this formalism is general, we carry out the numerical demonstration for a typical open quantum system--the spin-boson model. 
\end{abstract}

\maketitle
Time correlation functions (TCFs) can be used to directly quantify dynamic quantities, such as transport coefficients~\cite{kubo1957statistical,kubo1957statistical2}, absorption spectra~\cite{mukamel1995principles}, and chemical rate constants~\cite{kubo1957statistical2,yamamoto1960quantum,voth1989time}, which captures the response of a system to weak perturbations~\cite{kubo1991nonequili}. Consequently, researchers have dedicated significant efforts to develop accurate and efficient methods for calculating TCFs, resulting in a wide range of analytical and computational approaches.

Calculating TCFs presents a challenge in finding a balance between computational feasibility and accuracy. Numerically exact schemes offer high accuracy but are limited to small, idealized models, and their computational cost tends to scale unfavorably or exponentially with simulation time~\cite{mak1990solving,mak1991coherent,egger1992quantum,egger1994low,shao2001iterative,shao2002iterative,tanimura2014reduced,tanimura2015real,song2015new}. Approximate methods, such as perturbative approaches~\cite{bloch1957generalized,redfield1965theory,aslangul1986spin,leggett1987dynamics}, analytic continuation~\cite{jarrell1996maximum,gallicchio1996calculation,boninsegni1996density,sim2001quantum,krilov2001quantum,golosov2003analytic}, quantum mode coupling theory~\cite{denny2001mode, rabani2002self,reichman2002self2,rabani2004fully,shental2004mode, reichman2005mode,rabani2005quantum,markland2011quantum}, \textcolor{red}{process tensor methods~\cite{fux2024oqupy,ortega2024unifying}} and quasi-~\cite{tully1971trajectory,gerber1982time,tully1990molecular,stock1995semiclassical,tully1998mixed} and semi-classical~\cite{meyera1979classical,cao1994formulation,stock1997semiclassical,wang1998semiclassical,sun1998semiclassical,miller1998direct,egorov1998semiclassical,egorov1999quantum,jang1999path,jang1999derivation,miller2001semiclassical,shi2003relationship,shi2003semiclassical,craig2004quantum,craig2005chemical,craig2005refined,liu2007real,miller2009electronically,habershon2013ring} schemes, are more scalable to realistic multidimensional systems, but suffer from limited accuracy and convergence issues. Consequently, despite the availability of various TCF calculation methods, there is a clear demand for the development of widely applicable, accurate, and computationally efficient approaches.

Efficient and accurate calculations of TCFs can be achieved by utilizing the Mori formalism, which is based on the projection operator technique~\cite{fick1990quantum,grabert2006projection}. The Mori approach provides a low-dimensional equation of motion, known as the generalized quantum master equation (GQME), for the TCF~\cite{zhang2006nonequilibrium,kelly2016generalized}. The reduced dimensionality of the GQME corresponds to the space spanned by the probed observables~\cite{zhang2006nonequilibrium}, while the influence of excluded degrees of freedom is captured in the memory kernel~\cite{zhang2006nonequilibrium,kidon2018memory}.
Computing the memory kernel involves dealing with two challenges: the full dimensionality of the original problem and the application of the projected propagator, $e^{i\bm{\mathcal{Q}}\mathcal{L}t}$~\cite{montoya2016approximate}. The former can be addressed using numerically exact~\cite{shi2003new,zhang2006nonequilibrium,cohen2011memory,cohen2013numerically,cohen2013generalized,wilner2013bistability,wilner2014nonequilibrium, rosenbach2016efficient,yan2016dissipation} or approximate methods~\cite{shi2004semiclassical,kelly2013efficient,kelly2015accurate,pfalzgraff2015nonadiabatic,montoya2016approximate,kelly2016generalized}. Studies have shown that the short lifetime of the memory kernels can significantly enhance computational efficiency, regardless of the specific method employed to calculate them~\cite{kidon2018memory}. The latter challenge can be easily handled by implementing the hierarchical equations of motion (HEOM)~\cite{tanimura1990nonperturbative,tanimura2006stochastic,yan2004hierarchical,xu2005exact,xu2007dynamics,jin2008exact} or its second quantization generalization, the dissipation equation of motion (DEOM)~\cite{yan2016dissipation}. However, the computation cost of the DEOM largely increases as the number of dissipatons grows~\cite{yan2016dissipation}.

In this Letter, we introduce a novel and efficient method, referred to as the memory kernel coupling theory (MKCT), for calculating TCFs solely based on time-independent moments. Our approach involves decomposing the TCF memory kernels into auxiliary kernels, which are coupled through the equations of motion and higher-order moments. As the number of decompositions increases, these auxiliary kernels rapidly decay, allowing us to truncate the equations of motion and obtain the memory kernel. The coupled equations of motion can be solved either in the time domain or frequency domain. Importantly, our method only requires a single steady-state calculation to construct the auxiliary kernels, making it scalable and independent of system size. We demonstrate that the MKCT provides results consistent with exact numerical methods such as the DEOM. Our method opens up new avenues for studying the dynamics in complex quantum systems.

    \begin{figure*}
    \centering
    \includegraphics[width=.98\textwidth]{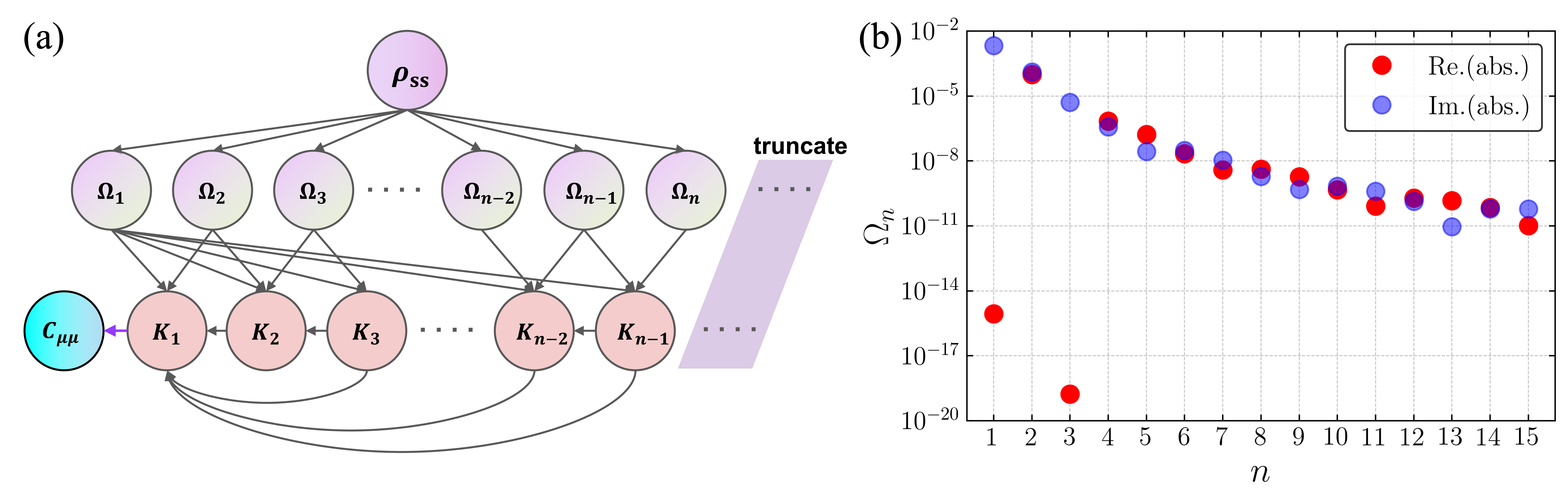}
    \caption{(a) A schematic diagram of MKCT. To calculate the correlation function $C_{\mu\mu}$, only the steady state $\rho_{\rm ss}$ is required. The symbols have been explained in the main text. (b) The absolute values of the real and imaginary parts of $\Omega_n$. $\Omega_n$ decays rapidly as $n$ increases, so we can truncate at an appropriate $n$ like FIG.~\ref{fig:sk} (a). Our parameters are $\epsilon = \Delta$, $k_B T=0.1\Delta$, $\gamma=\Delta$, and $\eta=0.1\Delta$.  We set $\hbar = 1$.}
    \label{fig:sk}
\end{figure*}

We start the theory by considering the auto-correlation function defined as 
\begin{equation}
C_{\mu\mu}(t) \equiv \braket{\hat{\mu}(t)\hat{\mu}(0)} = \text{Tr}[\hat{\mu}e^{-i\mathcal{L}t}(\hat{\mu}\rho_{\rm ss})],
\label{eqn:CORR}
\end{equation}
with $\hat\mu$ being an arbitrary Hermitian operator. Here, the dynamic propagator is generated by the Liouvillian $\mathcal L(\cdot)\equiv[H, (\cdot)]$, with $H$ being the Hamiltonian, and $\rho_{\rm ss}$ is the corresponding steady-state density operator.  For open quantum systems, we can employ the self-consistent iteration algorithm (SCI)~\cite{zhang2011note} to accurately and efficiently evaluate the steady-state, $\rho_{\rm ss}$.

To obtain the Mori-based GQME, we define the following projection operator $\bm{\mathcal{P}}$, 
\begin{equation}
    \bm{\mathcal{P}}\hat{A}=\braket{\hat{A}\hat{\mu}}\hat{\mu}/\braket{\hat{\mu}\hat{\mu}},
\end{equation}
where the inner product is given by 
\begin{equation}
    \braket{\hat{A}\hat{B}} \equiv \text{Tr}(\hat{A}\hat{B}\rho_{\rm ss})
\end{equation}
with $\hat A$ and $\hat B$ being arbitrary Hermitian operators. According to the Mori-Zwanzig formalism~\cite{nakajima1958quantum,zwanzig1960ensemble,mori1965transport}, we have
\begin{equation}
    \dot{\hat{\mu}}(t) = \Omega_{1}\hat{\mu}(t) + \int_{0}^{t}\! d\tau\, K_{1}(\tau)\hat{\mu}(t-\tau)+\hat{f}(t).
\end{equation}
Here, we have defined the higher-order moments 
\begin{equation}
    \Omega_{n} \equiv \braket{(i\mathcal{L}_{\text{}})^{n}\hat{\mu}\hat{\mu}}/\braket{\hat{\mu}\hat{\mu}}
    \label{eqn:Omega_n}
\end{equation}
and the auxiliary kernels 
\begin{equation}
    K_{n}(t) \equiv \braket{(i\mathcal{L})^{n}\hat{f}(t)\hat{\mu}}/\braket{\hat{\mu}\hat{\mu}}.
    \label{eqn:Kn}
\end{equation}
$\hat{f}(t)$ in the above equation is referred to as the random force operator
\begin{equation}
    \hat{f}(t) \equiv e^{it\bm{\mathcal{Q}}\mathcal{L}_{}}\bm{\mathcal{Q}}i\mathcal{L}_{\text{}}\hat{\mu},
    \label{eqn:f}
\end{equation}
with $\bm{\mathcal{Q}} = \bm{\mathcal{I}} - \bm{\mathcal{P}}$ being the complementary projection operator. 

Thus,  we have the derivative of Eq.~\ref{eqn:CORR}:~\cite{montoya2016approximate,bhattacharyya2024mori}
\begin{equation}
    \dot{C}_{\mu\mu}(t) = \Omega_{1} C_{\mu\mu}(t) + \int_{0}^{t} d\tau K_{1}(\tau)C_{\mu\mu}(t-\tau) + D(t).
    \label{eqn:CORR_dot}
\end{equation}
Here, the inhomogeneous term $D(t)$ can be removed via the proper choice of projection operator and is zero for us. Please refer to Supplemental Material (SM) for details.
Now if we want to get $C_{\mu\mu}(t)$, the only thing we need to know is the $K_1(t)$. However, as mentioned at the beginning of the Letter, it is difficult to directly calculate $K_1(t)$ according to Eq.~\ref{eqn:Kn}, as the calculation of $\hat{f}(t)$ (Eq.~\ref{eqn:f}) involves the application of the projected propagator. So next, we introduce the most crucial part of the Letter, where we decompose $K_1$ to avoid time evolution at the quantum mechanical level.

Firstly, $K_1(0)$ can be directly obtained without time evolution based on Eq.~\ref{eqn:Kn} using $\hat{f}(0)=\bm{\mathcal{Q}}i\mathcal{L}_{\text{}}\hat{\mu}$ in Eq.~\ref{eqn:f}. More generally, one can show 
\begin{equation} \label{eq:kn0}
    K_{n}(0) = \Omega_{n+1} - \Omega_{n}\Omega_{1}.
\end{equation}
Therefore, we only need to consider $\dot{K}_{1}(t)$, which can be obtained directly:
\begin{equation}
\begin{aligned}
    \dot{K}_{1}(t) &= \frac{(i\mathcal{L}_{}\dot{\hat{f}}(t),\hat{\mu})}{(\hat{\mu},\hat{\mu})}
    = K_{2}(t)-\Omega_{1}K_{1}(t).
\end{aligned}
\end{equation}
Similarly, we can show that the auxiliary kernels are coupled through
\begin{equation}
\begin{aligned}
    \dot{K}_{n}(t) &= K_{n+1}(t)-\Omega_{n}K_1(t).
    \label{eqn:Kndot}
\end{aligned}
\end{equation}
Eq.~\ref{eqn:Kndot} is one of the key findings in the Letter, which presents the coupled equations of motion for the auxiliary kernels. Hence, we term our formalism as memory kernel coupling theory (MKCT). With the proper truncation in the order of $n$ and initial conditions from Eq. \ref{eq:kn0}, we can solve the coupled equations of motion. Note that, even though the coupled equations of motion for the auxiliary kernels look similar to the HEOM for the auxiliary densities, the structure is much simpler--the structure for auxiliary kernels is chain-like, whereas the structure for auxiliary density in HEOM is tree-like. See Fig.\ref{fig:sk} (a). Note also that $K_n(t)$ and $\Omega_n$ are numbers instead of matrices (or operators), such that the computational cost of solving the coupled equations of motion for auxiliary kernels is independent of system size.   
\begin{figure*}[ht]
    \centering
    \includegraphics[width=.90\textwidth]{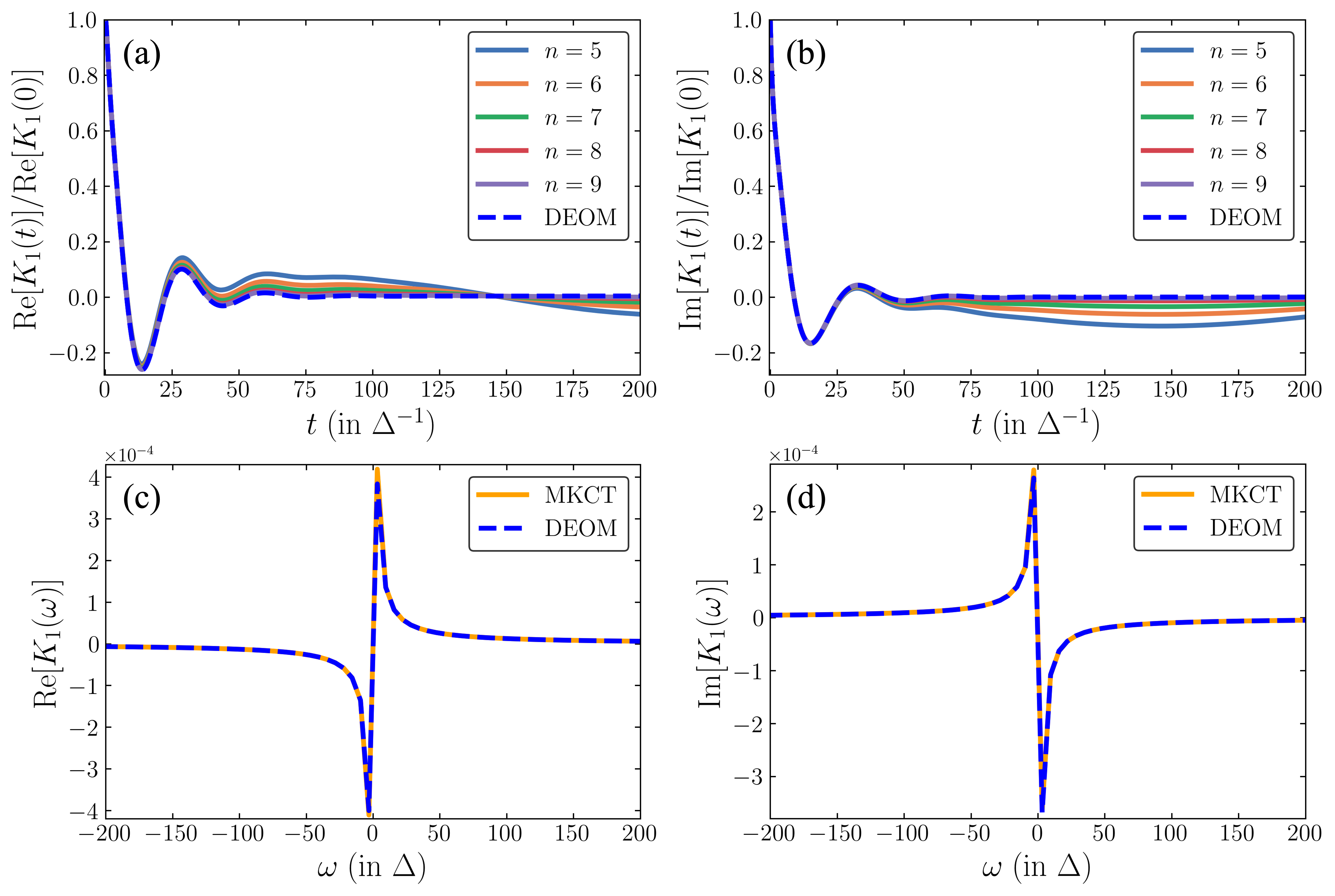}
    \caption{(a,b) The real and imaginary parts of the time-domain memory kernel $K_1(t)$. Here, $n$ represents the number of coupled equations of motion kept in the calculation. The blue dashed line is the direct calculation by DEOM through time evolution. (c,d) The real and imaginary parts of the frequency domain memory kernel $K_1(\omega)$. MKCT truncates at $n=9$. All other parameters are identical to FIG.~\ref{fig:sk}.}
    \label{fig:K_1}
\end{figure*}
 
The coupled equations of motion for the auxiliary kernels can be also solved in frequency domain, where we can apply Fourier transformation to Eq.~\ref{eqn:Kndot}:
\begin{equation}
    i \omega K_n(\omega) -K_n(0) = K_{n+1}(\omega)-\Omega_n K_1(\omega).
\end{equation}
Then, we can get $K_1$ in the frequency domain $\omega$ with proper truncation, such as $K_{n+1}(t) \approx 0 $:
\begin{equation}
\begin{aligned}
    K_1(\omega) = 
    \frac{\sum_{k=1}^{n}(i\omega)^{n-k}(\Omega_{k+1}-\Omega_{k}\Omega_{1})}{\sum_{k=0}^{n-1}(i\omega)^{n-k}\Omega_{k}}.
    \label{eqn:Kz}
\end{aligned}
\end{equation}
We have used the fact that $K_n(0)$ depends only on $\Omega_n$ from Eq. \ref{eq:kn0}. Thus, it is clear from the above equation that the memory kernel is completely captured by higher-order moments only $\Omega_n$. 



To show the applicability of the approach, we consider an open quantum system coupled to a Bose environment (bath), described by the Hamiltonian $H_{\text{}}=H_{\text{S}}+H_{\text{SB}}+h_{\text{B}}=H_{\text{S}}+ \hat{Q}\hat{F} + h_{\text{B}}$,
where $H_{\text{S}}$ and $h_{\text{B}}$ are the system Hamiltonian and the bath Hamiltonian, and
$H_{\text{SB}}$ is the coupling between the system and bath. $\hat{Q}$ is \textcolor{red}{a} system operator and $\hat{F}$ is \textcolor{red}{an} environment operator, which are both Hermitian. We have 
\begin{equation}
\begin{aligned}
        H_{\text{S}} &= \frac{1}{2}\Delta\sigma_z + \epsilon\sigma_x,\\
        h_{\text{B}} &= \sum_{j} \frac{1}{2} \omega_j (\hat{p}^2_j+\hat{x}^2_j),\\
        \hat{F} &= \sum_{j}c_j\hat{x}_j.
\end{aligned}  
\end{equation}
Here, $\Delta$ is the unit energy, which corresponds to the energy difference between the two sites. $\epsilon$, represents the tunneling matrix element, which is assumed to be static. And $\sigma_i$ corresponds to the $i^{\text{th}}$ Pauli matrix. $p_j$, $x_j$, and $\omega_j$ are the momenta, coordinates, and frequency for the $j^{\text{th}}$ harmonic oscillator, respectively. $c_j$ is the coupling constant that describes the strength of the interaction between the system and the $j^{\text{th}}$ oscillator.

The system-bath interaction is fully characterized by the spectral density,
\begin{equation}
    J(\omega) = \frac{\pi}{2}\sum_j c_j^2\delta(\omega-\omega_j) = \frac{2\eta\gamma\omega}{\omega^2 + \gamma^2},
    \label{eqn:Drude}
\end{equation}
which encodes the frequency-resolved coupling between the system and the oscillators that compose the bath. The second equal sign in Eq.~\ref{eqn:Drude} corresponds to the often used Drude form for the spectral density~\cite{weiss2012quantum,kleinert2009path,yan2005quantum}.

As shown in FIG.~\ref{fig:sk} (a), we show a schematic diagram of our theory. If $\Omega_n$ decays to near $0$ with the increase of $n$, we can truncate Eq.~\ref{eqn:Kndot} to obtain $\dot{K}_{1}(t)$. In the end, $C_{\mu\mu}$ depends only on $\rho_{\rm ss}$ through $K_1$ and $\Omega_1$. 

\begin{figure}[h]
    \includegraphics[width=.45\textwidth]{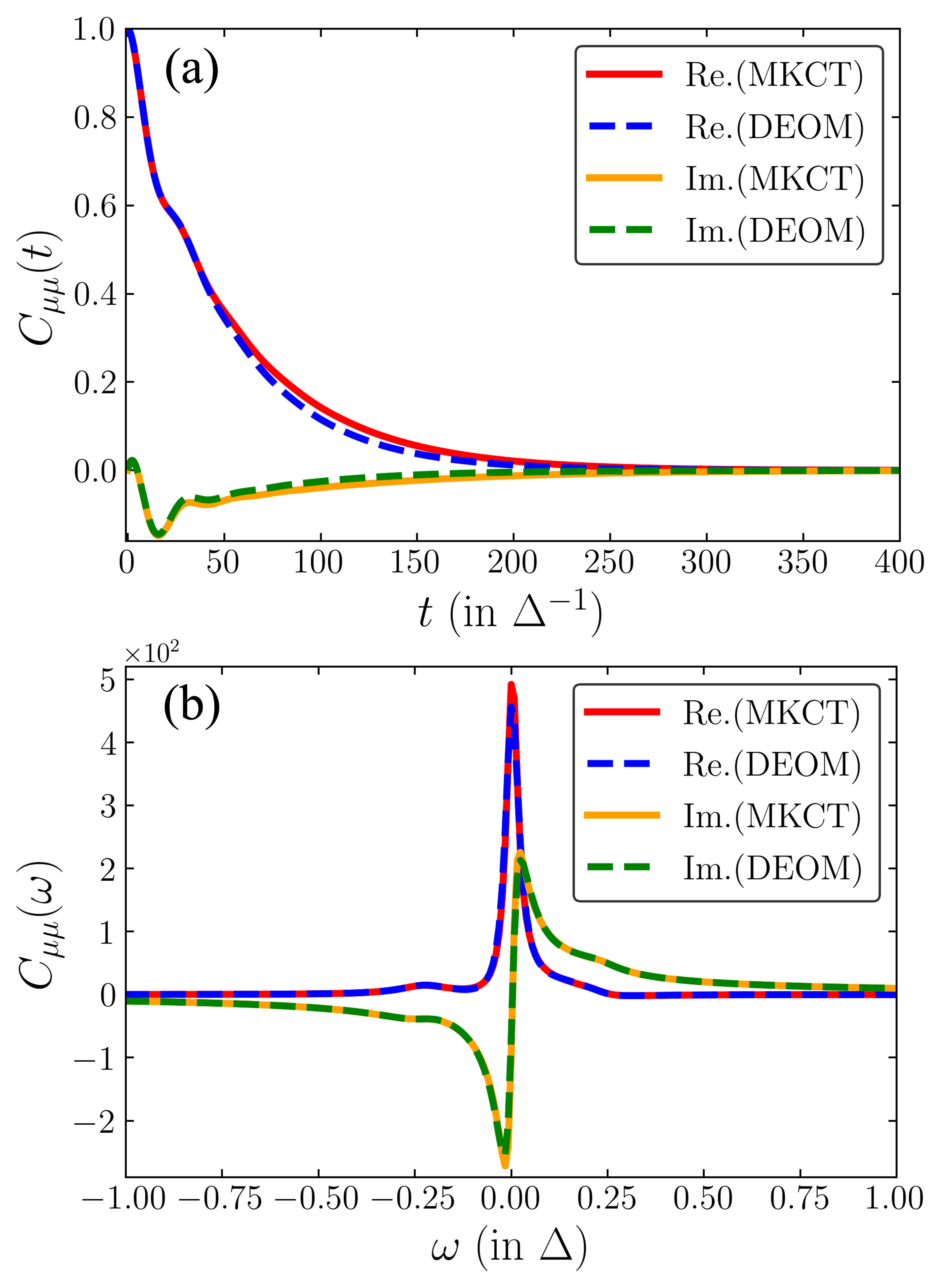}
    \caption{(a) The real and imaginary parts of the time-domain correlation function $C_{\mu\mu}(t)$. MKCT truncates at $n=9$. (b) The real and imaginary parts of the frequency domain correlation function $C_{\mu\mu}(\omega)$ calculated from (a). All other parameters are identical to FIG.~\ref{fig:sk}.}
    \label{fig:CORR}
\end{figure}

We show the results calculated by the approach introduced above with $\hat{\mu}(0)=\sigma_x$. 
In FIG.~\ref{fig:sk} (b), we show the absolute values of the real and imaginary parts of $\Omega_n$ calculated according to Eq.~\ref{eqn:Omega_n}. When $n$ increases from $1$ to $15$, $\Omega_{n}$ decays from the order of $10^{-2}$ to the order of $10^{-10}$. This kind of decay of $\Omega_n$ ensures the possibility of truncation. As shown in FIG.~\ref{fig:K_1} (a,b), we employ various truncations, namely $n=5,6,...,9$, to compute the memory kernel $K_1(t)$. We can see that both the real and imaginary parts of $K_1(t)$ will gradually approach the accurate value calculated by DEOM as the truncation position $n$ increases. In FIG.~\ref{fig:K_1} (c,d), we use Eq.~\ref{eqn:Kz} to get the real and imaginary parts of $K_1$ with $n=9$. The results are both in line with the Fourier transformation of $K_1(t)$ of DEOM as shown in FIG.~\ref{fig:K_1} (a,b). 
In FIG.~\ref{fig:CORR}, we use $K_1(t)$ with $n=9$ to calculate $C_{\mu\mu}$ according to Eqs.~\ref{eqn:CORR} and ~\ref{eqn:CORR_dot}. These time and frequency domain results are all in line with the results that were directly calculated by DEOM. Compared with Fig.~\ref{fig:K_1} (a,b) and FIG.~\ref{fig:CORR} (a), we observe that the memory kernel $K_1(t)$ decays much faster than the correlation function $C_{\mu\mu}(t)$. This finding further supports the notion of using the memory kernel to accelerate the calculation of the TCFs.
However, there are still cases where the memory kernel does not decay fast enough, such that longer dynamics are required to compute the TCFs~\cite{denny2001mode, shental2004mode, reichman2005mode}. Our theory can be very helpful in that case, since the complexity of our formalism is independent of time (see Eq.~\ref{eqn:Kz}). 

\textcolor{red}{To demonstrate the generality and efficiency of MKCT, we also performed simulations on the single impurity Anderson model (SIAM)~\cite{liang2002kondo,farinacci2020interfering,moca2021kondo,kurzmann2021kondo,ferrier2020quantum} as shown in Fig.~\ref{fig:Fermi} (details in SM). SIAM is frequently exploited to study the Kondo resonance, which is computationally expensive since it manifests strong electronic correlations at very low temperatures. The system of SIAM is $H_{\text{S}}=\xi (\hat{n}_{\uparrow}+\hat{n}_{\downarrow}) + U\hat{n}_{\uparrow}\hat{n}_{\downarrow}$. The environment is $h_{\text{B}}=\sum_{ks} \xi_{ks}\hat{d}_{ks}^{\dagger}\hat{d}_{ks}$, which is composed of free electrons. The interaction between the system and the environment is described by Lorentzian spectral density,
\begin{equation}
    J_{\text{L}}(\omega)=\frac{\Delta W^2}{\omega^2+W^2}.
\end{equation}
The simulation results are exhibited with the time correlation function $A_s(t) = \braket{\{\hat{a}_{s}(t),\hat{a}_{s}^{\dagger}(0)\}}$ and impurity spectral density,
\begin{equation}
    A_{s}(\omega) = \frac{1}{2\pi}\int_{-\infty}^{\infty} \text{d}t \; e^{i\omega t} \braket{\{\hat{a}_{s}(t),\hat{a}_{s}^{\dagger}(0)\}},
\end{equation}
where $\hat{a}_{s}^{\dagger}\;(\hat{a}_{s})$ is the creation (annihilation) operator of the electron with spin $s$ and $\{,\}$ denotes the anti-commutator. 
First, in Fig.~\ref{fig:Fermi} (a), we see that MKCT exactly captures $A_s(t)$ with a truncation $n=16$ as  compared to DEOM. As side note, the differences appearing in Fig.~\ref{fig:CORR} (a) can be removed by improving $n$. Next, in Fig.~\ref{fig:Fermi} (b), note that the Kondo resonance $\pi\Delta A(0)$ and the Fermi resonance that occur at $\omega= \pm U/2$ are also exactly calculated by MKCT. Finally, in Fig.~\ref{fig:Fermi} (c), we show that MKCT is more than $28$ times faster than DEOM in calculating the SIAM, and the marginal cost of MKCT calculations is lower compared to DEOM.}

In conclusion, we present a general formalism, termed the memory kernel coupling theory (MKCT), to calculate TCFs. We decompose the memory kernels of TCFs into auxiliary kernels, such that we can construct a chain-like coupled equations of motion. The coupled equations of motion for the auxiliary kernels can be solved with proper truncation, similar to the HEOM. Notably, our method requires only a single steady-state calculation to construct the auxiliary kernels, such that the complexity of solving the coupled equations of motion is independent of the system size. We also present the solution to the coupled equations of motion in the frequency domain. 

Additionally, we demonstrate that our method can reproduce results from exact numerical methods for two typical open quantum systems--the spin-boson model and the SIAM. We notice that memory kernels can exhibit shorter lifetimes compared to the lifetime of the TCFs. In the future, we aim to extend the auxiliary kernel framework to multiple-time correlation functions. Note also that our method is not restricted to the case with clear system-bath separation, or bath with Gaussian statistics. Indeed, MKCT can be applied to the systems that are not easily handled by HEOM or DEOM.

Moreover, our formalism can be combined with other methods to solve the dynamical problems of complex quantum systems. For instance, in quantum chemistry, the ground state properties can be obtained using variational principles. Using the ground-state moments, one can calculate TCFs using our method, thereby obtaining information about excited states. Similarly, obtaining dynamic properties directly from Quantum Monte Carlo (QMC) is challenging due to the sign problem. The MKCT can utilize ground-state properties obtained from QMC to calculate dynamic properties. Such that we expect our formalism can have a wide range of applicability. 

\onecolumngrid
\begin{center}
\begin{figure}
    \centering
    \includegraphics[width=.98\textwidth]{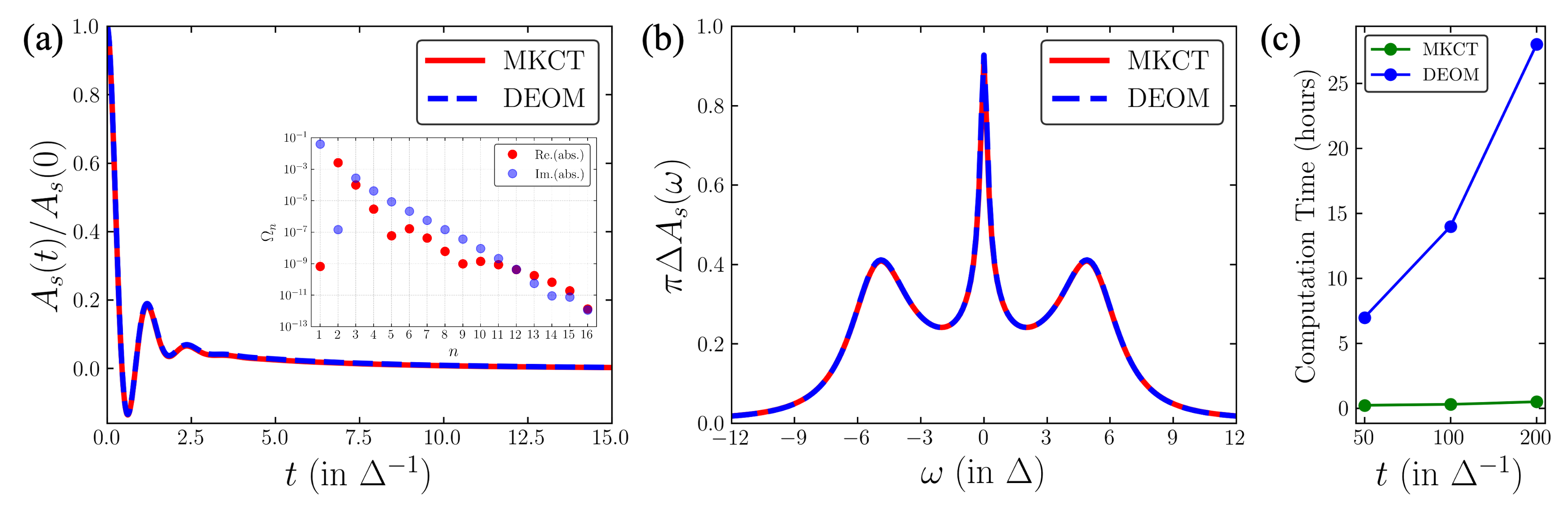}
    \caption{\textcolor{red}{(a) Moments $\Omega_n$ of SIAM and the corresponding time-domain correlation function $A_{s}(t)$. (b) The frequency-domain correlation function $A_{s}(\omega)$ of SIAM. (c) Comparison of SIAM computation time between MKCT and DEOM on AMD EPYC 7502 (2.5 GHz, 32 cores) with increasing SIAM simulation time $t$. In these simulations, the number of exponential terms is $8$ and we set the DEOM truncation tier to be $6$. MKCT truncates at $n=16$. Our parameters are $k_{\text{B}}T=0.01\Delta$, $W=10\Delta$, $U=8\Delta$, $\xi=-U/2$, $\text{d}t=0.001\Delta^{-1}$. We set $\hbar=1$.}}
    \label{fig:Fermi}
\end{figure}
\twocolumngrid
\end{center}


W.D. acknowledges the support from National Natural Science Foundation of China (No. 22361142829 and No. 22273075) and Zhejiang Provincial Natural Science Foundation (No. XHD24B0301). Y.W. thanks the support from the National Natural Science Foundation of China (Nos. 22103073 and 22373091). 


\bibliography{ref}

\end{document}